\documentclass[journal]{IEEEtran}

\usepackage[utf8]{inputenc}
\usepackage{graphicx}
\usepackage{caption}
\usepackage{subcaption}
\usepackage{amssymb}
\usepackage{color}
\usepackage{acronym}
\usepackage{amsmath}
\usepackage{algorithmic}
\usepackage{float}
\usepackage[square,sort,comma,numbers]{natbib}
\usepackage[linesnumbered,ruled,vlined]{algorithm2e}

\acrodef{vlc}[VLC]{visible light communication}
\acrodef{lifi}[LiFi]{light fidelity}
\acrodef{noma}[NOMA]{Non-Orthogonal Multiple Access}
\acrodef{ofdma}[OFDMA]{orthogonal frequency division multiple access}
\acrodef{tdma}[TDMA]{time division multiple access}
\acrodef{cdma}[CDMA]{code division multiple access}
\acrodef{oma}[OMA]{orthogonal multiple access}
\acrodef{irs}[IRS]{intelligent reflecting surface}
\acrodef{los}[LoS]{line-of-sight}
\acrodef{nlos}[NLoS]{non-line-of-sight}
\acrodef{led}[LED]{light-emitting diode}
\acrodef{fov}[FoV]{field-of-view}
\acrodef{ap}[AP]{access point}
\acrodef{re}[RE]{reflecting element}
\acrodef{sic}[SIC]{successive interference cancellation}
\acrodef{snr}[SNR]{signal-to-noise-ratio}
\acrodef{sinr}[SINR]{signal-to-interference-plus-noise-ratio}
\acrodef{pep}[PEP]{pairwise error probability}
\acrodef{ber}[BER]{bit-error-rate}
\acrodef{ga}[GA]{genetic algorithm}
\acrodef{mems}[MEMS]{micro-electromechanical systems}
\acrodef{fpga}[FPGA]{field-programmable gate array}
\acrodef{fso}[FSO]{free space optical}
\acrodef{owc}[OWC]{optical wireless communication}
\acrodef{rf}[RF]{radio frequency}
\acrodef{em}[EM]{electromagnetic}
\acrodef{ue}[UE]{user equipment}
\acrodef{oma}[OMA]{orthogonal multiple access}
\acrodef{noma}[NOMA]{non-orthogonal multiple access}
\acrodef{csi}[CSI]{channel state information}
\acrodef{pd}[PD]{photo-detector}
\acrodef{es}[ES]{exhaustive search} 
\acrodef{fpa}[FPA]{fixed power allocation} 
\acrodef{pls}[PLS]{physical layer security} 
\acrodef{awgn}[AWGN]{additive white Gaussian  noise} 
\acrodef{np}[NP]{ non-deterministic polynomial} 
\begin{document}
\title{Intelligent Reflecting Surfaces for Enhanced Physical Layer Security in  NOMA VLC Systems}

\author{{Hanaa~Abumarshoud,~Cheng~Chen,~Iman~Tavakkolnia,~Harald~Haas, and~Muhammad~Ali~Imran.}

\thanks{Hanaa~Abumarshoud~and~Muhammad~Ali~Imran are with James Watt School of Engineering, University of Glasgow, Glasgow, UK (e-mail:\{hanaa.abumarshoud, Muhammad.Imran\}@glasgow.ac.uk). }

\thanks{~Cheng~Chen,~Iman~Tavakkolnia,~and~Harald~Haas are with the LiFi R$\&$D Centre, University of Strathclyde, Glasgow, UK  (e-mail:\{c.chen,i.tavakkolnia, harald.haas\}@strath.ac.uk). }}

\maketitle

\begin{abstract}
The rise of intelligent reflecting surfaces (IRSs)  is opening the door for unprecedented capabilities in visible light communication (VLC) systems. By  controlling light propagation in indoor environments, it is possible to manipulate the channel conditions  to achieve specific key performance indicators. In this paper, we investigate the role that IRSs can play in boosting the secrecy capacity of non-orthogonal multiple access (NOMA) VLC systems. More specifically, we propose an IRS-based physical layer security (PLS) mechanism that mitigates the information leakage risk inherent in NOMA. Our results demonstrate that the achieved secrecy capacity can be enhanced by up to  $105\%$ for a number of $80$ IRS  elements. To the best of our knowledge, this is the first paper that examines the PLS of NOMA-based IRS-assisted VLC systems. 
\end{abstract}

\section{Introduction}
\Ac{vlc} has  gained significant interest in the last decade, which has been manifested by extensive research studies as well as practical  implementations from academia and industry alike. Based on \ac{vlc}, the concept of \ac{lifi} started to take shape as a fully networked solution for bi-directional high-speed connectivity with   multiple  access support \cite{HAAS2020443,8932632}. 
Different optical multiple access schemes  have been investigated in order to facilitate  ubiquitous multi-user connectivity in \ac{lifi}. Multiple access can be broadly divided into two categories: \ac{oma}  and \ac{noma}. \ac{oma} allocates  orthogonal resources to the users either in  the frequency domain, by means of \ac{ofdma}, or the time domain, by means of \ac{tdma}. \ac{noma} schemes enable  multiple users of accessing   the same  bandwidth and time resources by allocating distinct power levels, in the case of power-domain \ac{noma}, or distinct codes, in the case of \ac{cdma} \cite{9137669, 8713381}. In this paper, we refer to power-domain \ac{noma} as \ac{noma}.

\ac{noma} is considered a promising solution for  boosting the  spectral efficiency of \ac{vlc}, mainly due to the inherent high \ac{snr} in indoor \ac{vlc} systems. To implement  \ac{noma},  superposition coding is performed  at the \ac{ap}. In this step,  distinct power levels are assigned to the different users' signals. The power allocation coefficients are  typically decided  based on the relative users' channel conditions such that users with higher channel gains are assigned  lower power levels, and vice versa.  At the receiver side, \ac{sic} is performed  to decode and subtract the signals with higher power levels first until the desired signal is extracted. 
\ac{sic} implies that users with higher decoding order, i.e., users allocated less power,  can successfully decode the signals of the users with lower decoding order. This presents a security gap in \ac{noma}-based communications if one of the network users is malicious. 

The concept of \acp{irs} has recently evolved as  a focal point of interest  in the wireless communication community as it offers a spectrum, energy, and cost-efficient approach for  sustainable evolution in wireless systems.
An \ac{irs} constitutes a number of \acp{re} that can be  artificially engineered in order to control their response to  incident  light rays.  Based on this, it is possible to effectively control the propagation of the light signals to achieve desired performance gains. A detailed  overview of the advantages and challenges related to the integration of \ac{irs} technology in the context of \ac{vlc} and \ac{lifi} systems is presented in \cite{abumarshoud2021}. Recent research efforts have considered  \ac{vlc} performance evaluation and enhancement  in \ac{irs}-enabled environments. For example, 
the energy efficiency maximisation of \ac{irs}-assisted \ac{vlc} systems was investigated  in \cite{9348585}  while  \cite{9526581} considered the optimisation of the \ac{irs} reflection coefficients with the objective of sum-rate maximisation. In \cite{9838853}, a framework for an \ac{irs}-assisted \ac{noma} \ac{vlc} system was presented with the objective of enhancing the link reliability.

In this paper, we propose a \ac{pls} mechanism for  \ac{irs}-assisted \ac{noma} \ac{vlc} systems. The term \ac{pls} refers to the  techniques  that  exploit the physical properties of the optical channel to secure the transmission of information against potential eavesdroppers \cite{9070153, pls, 9524909}. Assuming that network users are assigned a trust score, i.e., based on their previous activity, we formulate a problem to maximise the secrecy capacity of a trusted user while maintaining minimum rate constraints for the untrusted user. This is achieved by jointly optimising the \ac{noma} power allocation and the \ac{irs} configuration  based on the given system parameters, users' locations, and rate requirements. We propose a novel \ac{pls} strategy and an alternating optimisation algorithm that utilises the  adaptive-restart \ac{ga} in order to obtain a  computationally efficient solution.  The rest of the paper is organised as follows: the system model and problem statement are presented in Section \ref{sec:model}. The proposed joint \ac{irs} configuration and power allocation scheme is discussed in Section \ref{sec:secure}. Simulation results are shown in Section \ref{sec:results} and the paper is concluded in Section \ref{sec:conclusion}.

\section{System Model} \label{sec:model}
\begin{figure}[ht]
	\centering
	\resizebox{0.9\linewidth}{!}{\includegraphics{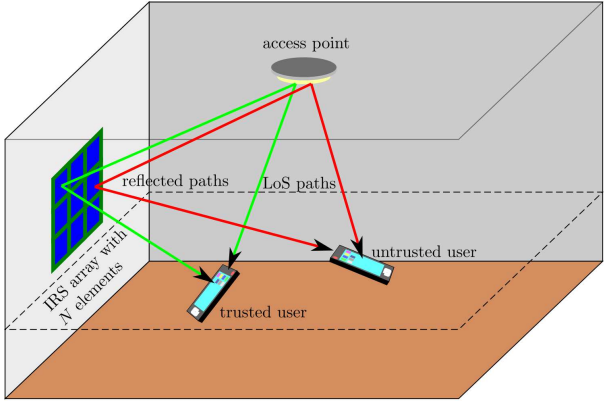}}
	\caption{System model with two \ac{noma} users: trusted and untrusted. The goal is to secure the transmission of the trusted user while satisfying given  rate constraints. }
	\label{fig:sys}
\end{figure}

We consider a \ac{noma}-based \ac{vlc} network consisting of one \ac{ap}, two paired users, and $N$ \acp{re}, as depicted in Fig. \ref{fig:sys}. It is noted that typical systems employing \ac{noma} divide users into multiple orthogonal pairs, i.e., in the time or frequency domain, and \ac{noma} is implemented within
each pair. As such, the users in this scenario may correspond to a \ac{noma} pair in a hybrid \ac{noma}/\ac{oma} network. 
 Without loss of generality, we assume that the user locations and \ac{vlc} \ac{csi} are available at a central control unit where the resource allocation decisions are made. 

\subsection{Channel gain of LoS paths}
The \ac{los} channel gain from the \ac{led} to the $k$-th user's \ac{pd} is given by
\begin{equation} \label{equ:h_los}
h_{k, {\rm LED}}^{\rm LoS} = \mathcal{V}_{k, {\rm LED}}^{\rm LoS} 
\cos^m(\phi_{k, {\rm LED}})    \cos(\psi_{k, {\rm LED}}), 
\end{equation} 
for $0 \leq \psi_{k, {\rm LED}} \leq \Psi_c$ and $0$ otherwise, where 
\begin{equation} \label{equ:h_los1}
\mathcal{V}_{k, {\rm LED}}^{\rm LoS}=\frac{A(m+1)}{2 \pi d_{k, {\rm LED}}^2} \mathcal{G}_f \mathcal{G}_c,    \end{equation}
where $d_{k, {\rm LED}}$ denotes  the Euclidean distance from the \ac{led} to the $k$-th user;  $A$ is the physical area of the \ac{pd};  $\phi_{k, {\rm LED}}$  is the angle of irradiance with respect to the axis normal to the \ac{led} plane; $\psi_{k, {\rm LED}}$ is  the angle of incidence with respect to the axis normal to the receiver plane. From analytical geometry, the irradiance and  incidence angles can be calculated as
\begin{subequations}
\begin{equation}
\cos(\phi_{k, {\rm LED}})  =\frac{{\bf{d}_{\rm \textit{k}, {\rm LED}}}\cdot{\bf{n}_{\rm{LED}}}}{\Vert {\bf{d}_{\rm \textit{k}, {\rm LED}}}\Vert },
\end{equation}
\begin{equation} \label{equ:cospsi1}
  \cos(\psi_{k, {\rm LED}})=\frac{{-\bf{d}_{\rm \textit{k}, {\rm LED}}}\cdot{\bf{n}_{\rm{PD}}}}{\Vert {\bf{d}_{\rm \textit{k}, {\rm LED}}}\Vert },
\end{equation}
\end{subequations}
where ${\bf{n}}_{\rm{LED}}$  and ${\bf{n}}_{\rm{PD}}$ are the normal vectors at the \ac{led} and the receiver planes, respectively, and the symbols $\cdot$ and $\Vert \cdot\Vert$ denote the inner product and the Euclidean norm operators, respectively. Furthermore, $\Psi_c$ denotes the \ac{fov} of the receiver; $\mathcal{G}_f$ is the gain of the optical filter;  $\mathcal{G}_c$ is the gain of the optical concentrator given by
\begin{equation}
\label{concentrator}
\mathcal{G}_c=\begin{cases}
\dfrac{\varsigma^2}{\sin^2\Psi_c}, & 0\le\psi\le\Psi_c\\
0, & {\rm otherwise}
\end{cases}, 
\end{equation}
where $\varsigma$ stands for the refractive index;  $m$ is the Lambertian order which is given by
\begin{equation}
\label{Lambertian}
m=-\frac{1}{\log_2(\cos\Phi_{1/2})},
\end{equation}
where $\Phi_{1/2}$ is the \ac{led} half-intensity angle.

\subsection{Channel gain of reflected paths via \ac{irs}}
\begin{figure*}
	\centering
	\resizebox{0.9\linewidth}{!}{\includegraphics{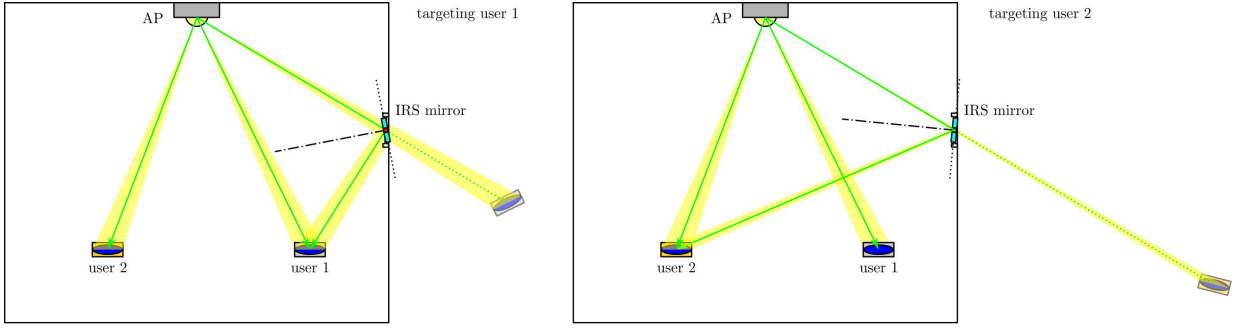}}
	  \caption{\ac{irs} configuration: each \ac{re} can serve one user at a time by steering the reflected light in a certain direction. } 
    \label{fig:IRS}
\end{figure*}

Optical \acp{irs} can be realised by means of metasurfaces or mirror arrays as discussed in \cite{9276478}. Different to \ac{rf} systems, where \acp{irs} can control the phase of the reflected signal, the focus in optical \acp{irs} is to steer the light. Phase control is not directly applicable to \ac{vlc} systems due to intensity modulation/ direct detection which deals only with the real part of the signal.  In this paper, we assume that the \ac{irs} is a mirror array consisting  of multiple adjacent  \acp{re}, each of which can be controlled to guide the impinging light towards a desired direction. Since light can be steered to specific point coordinates, each \ac{re} can only serve one user at a time, as seen in Fig. \ref{fig:IRS}.  
The specular reflection path can be considered as an alternative  path emitted from the \ac{ap} to the user. According to Snell's law of reflection, the orientation of each \ac{re} needs to be configured such that it reflects the impinging light towards a specific user. Based on that, each \ac{irs} \ac{re} can serve a single \ac{pd} at a certain time slot, and the interference from specular reflection paths is negligible.

Each reflected path through the \ac{irs} is composed of two components: \ac{led}-to-\ac{re} path and \ac{re}-to-\ac{pd} path. An approximate expression for the cascaded channel gain under the point source assumption was derived in \cite{9276478}. Based on this,  the cascaded channel gain through the $n$-th \ac{re} is given by 
\begin{equation} \label{equ:h_ref}
h_{k,n, {\rm LED}}^{\rm Ref} =  \mathcal{V}_{k,n, {\rm LED}}^{\rm Ref}   \cos^m(\phi_{n,{\rm{LED}}})    \cos(\psi_{k,n}), 
\end{equation}
for  $0 \leq \psi_{k,n} \leq \Psi_c$ and $0$ otherwise, where 
\begin{equation}  \label{equ:h_ref1}
\mathcal{V}_{k,n, {\rm LED}}^{\rm Ref}= \frac{A(m+1) }{2 \pi ({d_{n,{\rm{LED}}} +d_{k,n} )}^2 } \mathcal{G}_f \mathcal{G}_c .   
\end{equation}
we assume the \acp{re} have unity reflectivity for simplicity. Also,    $d_{n,{\rm{LED}}}$ and $d_{k,n}$ denote the Euclidean distance from the \ac{led} to the $n$-th \ac{re} and the $n$-th  \ac{re}  to the $k$-th user, respectively; $\phi_{n,{\rm{LED}}}$ and $\psi_{k,n}$ are the angle of irradiance and incidence with respect to the axis normal to the receiver plane, calculated as
\begin{subequations}
\begin{equation}
  \cos(\phi_{n,{\rm{LED}}})=\frac{\bf{d}_{\rm \textit{n},{\rm{LED}}}\cdot{\bf{n}_{\rm{LED}}}}{\Vert {\bf{d}_{\rm \textit{n},{\rm{LED}}}}\Vert },
\end{equation}
\begin{equation} \label{equ:cospsi2}
  \cos(\psi_{k,n})=\frac{\bf{d}_{\rm \textit{k},\rm \textit{n}}\cdot{\bf{n}_{\rm{PD}}}}{\Vert {\bf{d}_{\rm \textit{k},\rm \textit{n}}}\Vert },
\end{equation}
\end{subequations}
we define the \ac{nlos} channel gain vector $\tilde{\mathbf{h}_k}=[h_{k,1, {\rm LED}}^{\rm Ref}, h_{k,2, {\rm LED}}^{\rm Ref}, \dots , h_{k,n, {\rm LED}}^{\rm Ref} ]^T$. Moreover, since each \ac{re} can serve one user at a time, we define a binary \ac{irs} allocation vector $\mathbf{g}_k=[g_{k,1}, g_{k,2}, \dots , g_{k,N} ]^T$ such that $g_{k,n}=1$ of the $n$th \ac{re} is serving the $k$-th user, and $g_{k,n}=0$ otherwise.

\subsection{NOMA and the associated information leakage risk} 
According to the principle of \ac{noma}, the \ac{ap} allocates a higher power level to the user with lower channel gain and vice versa. The two signals are then superimposed in the power domain and transmitted in a single time/frequency resource block. On the receiver side, the user allocated higher power value can directly decode its intended signal while regarding the interference from the other user's signal as noise. However, the user allocated  low power level cannot directly decode its signal and needs to resort to \ac{sic}. In \ac{sic}, the user decodes and subtracts the  information sequence  intended for the other user to reach its intended information.  
The process of \ac{sic} necessities that one user decodes the other user's signal, which involves a security threat if that user was malicious. For this reason, we assume that each user is associated with a trust score, which can be obtained based on previous activity or through authentication measures. The user with a lower trust score is treated as a malicious user. In order to reduce the eavesdropping probability, the \ac{ap} needs to allocate a higher power level to the untrusted user (so that it cannot perform \ac{sic} and extract the confidential message). Implementing this power allocation strategy might be challenging in conventional \ac{noma}-based \ac{vlc} systems that rely merely on the \ac{los} channel paths. This is because the allocated power levels usually depend on  the associated channel gain, i.e., higher power is allocated to users with unfavourable channel conditions and vice versa. This challenge can be overcome in \ac{irs}-assisted \ac{vlc}. By configuring the allocation of the \ac{irs} \acp{re} between the two users to create better conditions for  enhanced \ac{pls}.

\section{Joint \ac{irs} Configuration and Power Allocation for Enhanced \ac{pls}} \label{sec:secure} 

This section proposes a novel  \ac{pls} approach, formulates a joint \ac{irs} configuration and power allocation problem to reduce the risk of eavesdropping, and proposes a solution based on an iterative optimisation algorithm.

\subsection{The proposed strategy}
Our goal is  to secure the signal of a trusted user, $\mathrm{U}_t$, against a potentially untrusted user, $\mathrm{U}_u$. Since both \ac{noma} users are network users, we need to satisfy the minimum rate constraints for both of them. This means that $\mathrm{U}_u$'s minimum rate requirment should be satisfied while, at the same time, minimising  its ability to decode the signal of $\mathrm{U}_t$. In the proposed strategy, the power allocation is not performed based on the users' channel gain values but rather based on the trust score, i.e., $\mathrm{U}_u$ is always in the first decoding order (always allocated higher power). In this way, $\mathrm{U}_u$ decodes its signal directly while $\mathrm{U}_t$ performs \ac{sic} to subtract the other users's signal and then decode its own signal.    According to the concept of \ac{noma}, the transmit signal is: 
\begin{equation}
x=P_t s_t + P_u s_u,     
\end{equation}
where $s_t$ and $s_u$ donate the information signal intended to $\mathrm{U}_t$ and $\mathrm{U}_u$, respectively;  $P_u > P_t$;  $P_u+P_t\leq P_{\rm LED}$. 
The received signals can be expressed as
\begin{subequations}
\begin{equation}
y_t= \rho \left(h_{t}^{\rm LoS} + \tilde{\mathbf{h}}_t^T   \mathbf{g}_t \right) \left(P_u s_u + P_t s_t\right)  + z_t,    
\end{equation}
\begin{equation}
y_u= \rho \left(h_{u}^{\rm LoS} + \tilde{\mathbf{h}}_u^T   \mathbf{g}_u \right) \left(P_u s_u + P_t s_t\right) + z_u,    \end{equation}
\end{subequations}
where $\rho$ is the \ac{pd} responsivity; $z_t$ and $z_u$ denote the receiver noise modeled as \ac{awgn} with variances $\sigma_t$ and  $\sigma_u$. Assuming that the trusted user performs successful \ac{sic}, the corresponding \ac{sinr} can be written as
\begin{subequations} 
\begin{equation}
\gamma_t=\frac{\rho^2 P_t^2    \left(h_{t}^{\rm LoS} + \tilde{\mathbf{h}}_t^T   \mathbf{g}_t \right)^2  }{\sigma_t^2} 
\end{equation}
\begin{equation}
\gamma_u=\frac{\rho^2 P_u^2 \left(h_{u}^{\rm LoS} + \tilde{\mathbf{h}}_u^T   \mathbf{g}_u \right)^2   }{P_t^2 \left(h_{u}^{\rm LoS} + \tilde{\mathbf{h}}_u^T   \mathbf{g}_u \right)^2 + \sigma_u^2}.
\end{equation}
\end{subequations}
The classic Shannon capacity formula does not accurately capture the data rates in \ac{vlc} systems, due to the distinct features of the optical channel as well as the illumination constraints. Alternatively, we utilise the tight lower bound in \cite{6636053} to calculate the achievable data rates at the  users as follows 
\begin{subequations}
\begin{equation}
R_t=\frac{1}{2} W \log_2 \left(1+\frac{e}{2\pi} \frac{\rho^2 P_t^2    \left(h_{t}^{\rm LoS} + \tilde{\mathbf{h}}_t^T   \mathbf{g}_t \right)^2  }{\sigma_t^2} \right)
\end{equation}
\begin{equation}
R_u=\frac{1}{2} W \log_2 \left(1+\frac{e}{2\pi} \frac{\rho^2 P_u^2 \left(h_{u}^{\rm LoS} + \tilde{\mathbf{h}}_u^T   \mathbf{g}_u \right)^2   }{P_t^2 \left(h_{u}^{\rm LoS} + \tilde{\mathbf{h}}_u^T   \mathbf{g}_u \right)^2 + \sigma_u^2}\right)
\end{equation}
\end{subequations}
where $W$ denotes the modulation bandwidth.

\subsection{Problem formulation}
Our objective is to maximise the secrecy capacity of $\mathrm{U}_t$, which can be defined by the tight lower bound in \cite{8421280} as  
\begin{equation}
C_t=\frac{1}{2} W \log_2 \left( \frac{\sigma_u^2}{2 \pi \sigma_t^2} \left(\frac{e \rho P_t^2 \left(h_{t}^{\rm LoS} + \tilde{\mathbf{h}}_t^T   \mathbf{g}_t \right)^2 + 2 \pi \sigma_t^2}{\rho P_u^2 \left(h_{u}^{\rm LoS} + \tilde{\mathbf{h}}_u^T   \mathbf{g}_u \right)^2 + \sigma_u^2 }\right)\right).  
\end{equation}
It is noted that, since $\mathrm{U}_u$ is an eligible network user, the optimisation approach  needs to ensure that the minimum rate requirement of $\mathrm{U}_u$ is satisfied. This implies that the resources allocated to $\mathrm{U}_u$ (allocated power and \acp{re}) facilitate that it can decode its own data signal, $s_u$, under the presence of interference from $\mathrm{U}_t$'s signal, without being able to extract $s_t$. The optimisation problem is formulated as 
\begin{subequations}
\begin{equation}
 \label{eq:of}
(\mathcal{P}): \; \underset{\mathbf{G}, \mathbf{P}}{\rm max}  \; C_t ,  
\end{equation}{\rm subject \;  to \;  } 
\begin{equation}\label{eq:c1}
 P_t + P_u \leq  P_{\rm LED} ,   
\end{equation}
\begin{equation} \label{eq:c2}
R_k \geq R^{\mathrm{min}}_k, \forall  k\in\{t,u\} 
\end{equation}
\begin{equation} \label{eq:c3}
g_{n,k}\in\{0,1\}, \forall n=1,2,\dots, N, \;  k\in\{t,u\} 
\end{equation}
\begin{equation} \label{eq:c4}
\sum_{k\in{t,u}}g_{n,k}\in\{0,1\}, \forall n=1,2,\dots, N. 
\end{equation}
\end{subequations}

Here, $\mathbf{P}=[P_t, P_u]$ is the \ac{noma} power allocation vector, and $\mathbf{G}=[\mathbf{g_t}, \mathbf{g}_u]$ is the \ac{irs} allocation matrix. Constraint (\ref{eq:c1}) is related to the total transmit power limitation, constraint (\ref{eq:c2}) ensures that the minimum rate requirements are satisfied, and constraints (\ref{eq:c3}) and (\ref{eq:c4}) ensure that each \ac{irs} \ac{re} can be allocated to a maximum of one user at a time. 
The optimisation problem  (\ref{eq:of}) constitutes  \ac{np}-hard problem. Next, we propose an iterative optimisation algorithm to reach a near-optimal solution. 
\subsection{Alternating optimisation algorithm}
In order to solve (\ref{eq:of}), we split the optimisation problem into two sub-problems, namely $\mathcal{P}_1$ and $\mathcal{P}_2$ that  are concerned with optimising a single variable, i.e., $\mathbf{G}$ and $\mathbf{P}$, respectively, while the other variable is fixed. The first sub-problem is the \ac{irs} \acp{re} allocation optimisation under fixed \ac{noma} power allocation vector $\mathbf{P}$, which can be expressed as 
\begin{subequations}
\begin{equation}
 \label{eq:p1}
(\mathcal{P}_1): \; \underset{\mathbf{G}}{\rm max}   \; C_t(\mathbf{G}) ,  
\end{equation} {\rm subject \;  to \;  } 
\begin{equation} \label{eq:p1c1}
R_k \geq R^{\mathrm{min}}_k, \forall  k\in\{t,u\} 
\end{equation}
\begin{equation} \label{eq:p1c2}
  0 \leq g_{n,k} \leq 1 \; \; \; \forall n=1,2,\dots, N, \;  k\in\{t,u\}
\end{equation}
\begin{equation} \label{eq:p1c3}
\sum_{k\in{t,u}}g_{n,k}\in\{0,1\}, \forall n=1,2,\dots, N. 
\end{equation}
\end{subequations}
Note that by relaxing the constraint in (\ref{eq:c3}) to  (\ref{eq:p1c2}), we obtain a linear programming problem that can be solved by means of the  CVX toolbox [32]. We then employ a greedy policy to recover the
discreteness of $\mathbf{G}$ as shown in steps  3-7 of Algorithm 1.
\begin{algorithm}[ht] 
\label{algo:1}
\SetAlgoLined
\KwIn{$h_{t}^{\rm LoS}$, $h_{u}^{\rm LoS}$, $\tilde{\mathbf{h}}_t$,$\tilde{\mathbf{h}}_u$, $R^{\mathrm{min}}_t$, $R^{\mathrm{min}}_u$, $\delta_1$.  }  

\KwOut{$\mathbf{G}$ and $\mathbf{P}$}  
\renewcommand{\labelenumi}{(\Roman{enumi})}  \bf{Init:} {\normalfont $i\leftarrow 0$, initialise $\mathbf{G}^{(0)}$ and $\mathbf{P}^{(0)}$  as random feasible solutions to $\mathcal{P}$.}\\ 
 \bf{repeat:} 
 {\normalfont 
Given $\mathbf{P}^{(i)}$  solve $\mathcal{P}_1$ to obtain \ac{irs} allocation matrix $\bf{G}^{(i)}$;
\\
\For{n=0:N} {
$k_\dag \leftarrow \underset{k}{\rm min} \; g_{n,{k_\dag}}$ ; \\ 
$g_{n,{k_\dag}} \leftarrow 1 $  and 
$g_{n,{k \neq  k_\dag}} \leftarrow 0 $;\\
$n \leftarrow n+1; $\\}
Given $\mathbf{G}^{(i+1)}$  solve $\mathcal{P}_2$ by \ac{ga} to obtain power allocation vector $\bf{P}^{(i+1)}$;\\
\bf{until:}   $|C_t(\mathbf{G}^{i+1},\mathbf{P}^{i+1})-C_t(\mathbf{G}^{i+1},\mathbf{P}^{i+1})|<\delta_1$ 
 }
\caption{Alternating optimisation algorithm}
\end{algorithm}

\begin{algorithm}[ht] 
\SetAlgoLined
\KwIn{Population size, $\mathcal{S}$, number of generations, $N_{\rm Gen}$, maximum run time, $t_{\rm max}$, $h_{t}^{\rm LoS}$, $h_{u}^{\rm LoS}$, $\tilde{\mathbf{h}}_t$,$\tilde{\mathbf{h}}_u$, $R^{\mathrm{min}}_t$, $R^{\mathrm{min}}_u$, $\mathbf{G}$.} 

\KwOut{Global best solution $\mathbf{P}$.}  
\bf{Start:} \\ 
{\normalfont Generate initial population of $\mathcal{S}$ chromosomes $\mathcal{Y}_0$ ;    \\
Set time counter $t=0$;}
 
 \While{$t<t_{\rm max}$}{ 
 \For{$i=1:N_{\rm Gen}$} {
 \For{$j=1:N_{\rm Gen}$} 
 { \normalfont
 Select a pair of chromosomes from $\mathcal{Y}_{j-1}$; \\ 
 Apply crossover operation on selected pair with crossover probability $\mathcal{P}_c$;  \\ 
 Apply mutation operation on the offspring with mutation probability $\mathcal{P}_m$;\\
 Check constraints \eqref{eq:p2c1}-\eqref{eq:p2c2} and repair; \\ 
 Evaluate the fitness of each  chromosome in $\mathcal{Y}_j$ using \eqref{eq:p2};  
  Select  elite chromosomes; }
 \normalfont Update $\mathcal{Y}_{\rm best}$; \\ 
 Restart with an adaptive initial population $\tilde{\mathcal{Y}}_(i+1)$ containing the elite chromosomes; 
 } }\label{algo:2}
\caption{Adaptive-restart genetic algorithm}
\end{algorithm}
\subsection {Adaptive-restart \ac{ga}}
Once the \ac{irs} allocation matrix $\mathbf{G}$ is obtained, we solve the \ac{noma} power allocation problem for fixed $\mathbf{G}$ as follows
\begin{subequations}
\begin{equation}
 \label{eq:p2}
(\mathcal{P}_2): \; \underset{\mathbf{P}}{\rm max}  \; C_t ,  
\end{equation}{\rm subject \;  to \;  } 
\begin{equation} \label{eq:p2c1}
R_k \geq R^{\mathrm{min}}_k, \forall  k\in\{t,u\} 
\end{equation}
\begin{equation}\label{eq:p2c2}
 P_t + P_u \leq  P_{\rm LED}.  
\end{equation}
\end{subequations}
In order to obtain a solution for $\mathcal{P}_2$, we utilise adaptive-restart \ac{ga} presented in Algorithm 2, which is known to be an effective  heuristic technique for obtaining computationally-efficient near-optimal solutions. Its working principle mimics  the idea of natural evolution based on  the idea of the "survival of the fittest".  
As shown in Algorithm 2, we start with an initial population of chromosomes, each representing a possible solution for $\mathbf{P}$. In each iteration of the algorithm, a mutation operation is  applied to the current  population in order to  generate slightly modified chromosomes  for the following generation, followed by a crossover operation that generates new instances of $\mathbf{P}$ within  the search space of candidate solutions.  The fitness of each candidate solution  is evaluated in each iteration based on (\ref{eq:p2}) and the associated constraints.  Based on that, a selection process is performed  to carry the fittest solutions to the next iteration and discard the bad solutions.  To enhance the global search capability of the \ac{ga} and reduce the probability of converging to a local optimum, We employ an adaptive-restart mechanism \cite{GHANNADIAN199681}. This is represented by the  outer \textit{for} loop in Algorithm 2 which restarts the \ac{ga} with some "elite" quality chromosomes generated in the previous iteration.

\begin{figure}[ht]
	\centering
	\resizebox{0.9\linewidth}{!}{\includegraphics{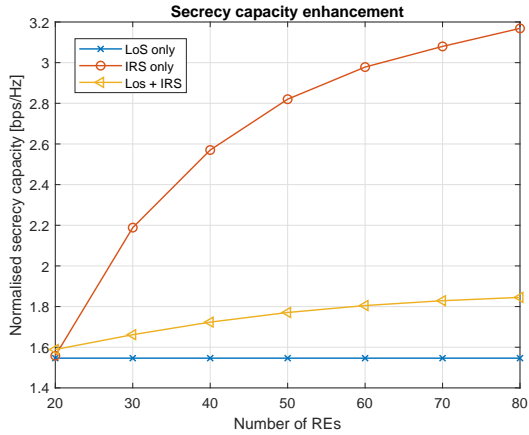}}
	\caption{The achievable secrecy capacity of U$_t$ vs the number of \acp{re} for transmit \ac{snr} of $80$ dB.}
	\label{fig:1}
\end{figure}
\begin{figure}[ht]
     \centering
     \begin{subfigure}[b]{0.9\linewidth}
        \centering
         \includegraphics[width=1\textwidth]{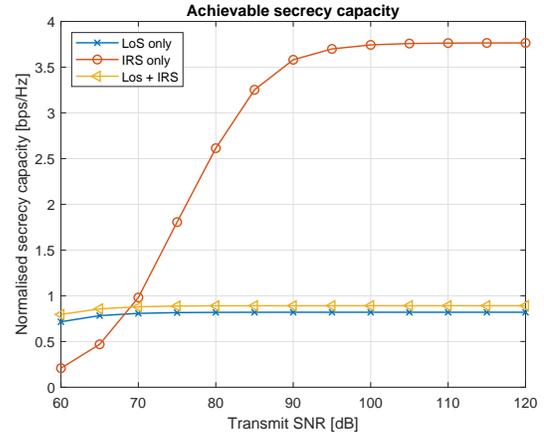}
         \caption{}
         \label{fig:a}
     \end{subfigure}
     \begin{subfigure}[b]{0.9\linewidth}
        \centering
         \includegraphics[width=1\textwidth]{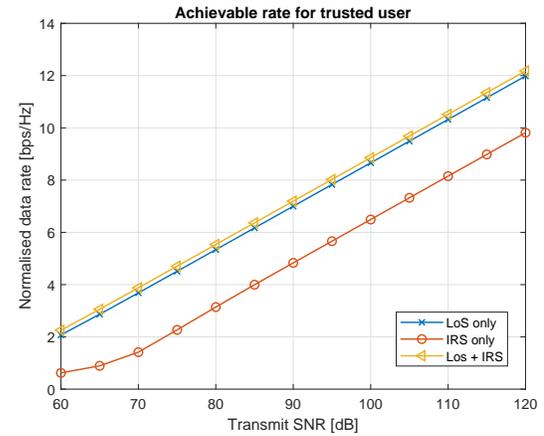}
         \caption{}
         \label{fig:b}
     \end{subfigure}
     \begin{subfigure}[b]{0.9\linewidth}
         \centering
         \includegraphics[width=1\textwidth]{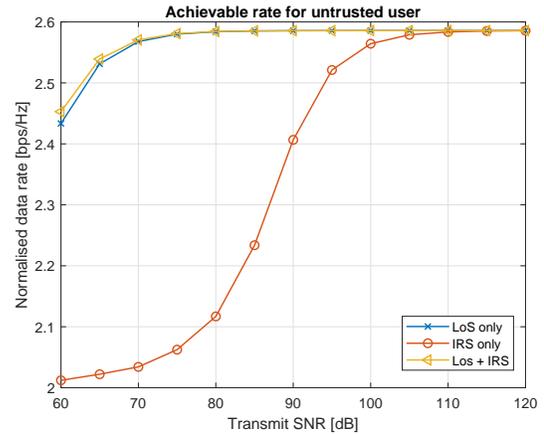}
         \caption{}
         \label{fig:c}
     \end{subfigure}
              \caption{The achievable performance in terms of (a) achievable
secrecy capacity, (b) achievable rate of U$_t$, and (c) achievable
rate of U$_u$, for an \ac{irs} fixed on one wall with 40 \acp{re}.}
     \label{fig:2}
\end{figure}
\section{Results and Discussions} \label{sec:results}

In this section, we present the simulation results for two \ac{noma} users located in a $3 \times 3 \times 5$ room. The transmitting \ac{led} \ac{ap} is fixed at the centre of the ceiling and the users' locations are randomly generated across the room for a number of $10^3$ iterations in order to capture the average system performance. We assume that an \ac{irs} consisting of $N$ \acp{re} is fixed on the wall. 
Moreover, we assume that the \ac{ap} has prior knowledge of the trust score of both users such that the user with a higher trust score is labeled as a trusted user, U$_t$, and the user with a lower trust score is labeled as an untrusted user, U$_u$. Based on the discussion in Section \ref{sec:secure},  U$_u$ is assigned the first \ac{sic} decoding order to reduce the risk of information leakage.

Fig. \ref{fig:1} presents the normalised secrecy capacity of U$_t$ under the proposed scheme vs the number of \acp{re}. It can be seen that using combined  \ac{los} and \ac{irs} links can slightly increase the secrecy capacity performance and that the enhancement becomes more pronounced as the number of \acp{re} increases. However, the performance enhancement can be considered limited and does not exceed $20\%$. This is attributed to the fact that the \ac{los} channel gain in \ac{vlc} remains dominant even for a high number of \acp{re}. Fig. \ref{fig:1} also shows the performance enhancement when only \ac{irs} links are utilised for communications. We can see that the secrecy capacity can be greatly enhanced with an improvement of up to $105\%$ for $80$ \acp{re}. This result might look confusing as the users experience much lower channel gain when there is no \ac{los} between the user and the \ac{ap}, Nonetheless, the performance enhancement here stems from the fact that the secrecy capacity is reliant on the relative channel gain values of the trusted users and the eavesdropper. In the absence of direct \ac{los} links, the \ac{irs} gives a great flexibility and a high degree of freedom to manipulate the users' channel conditions so as to maximise the achievable secrecy capacity.

In order to get more insights on the rate performance of the \ac{noma} users under the proposed scheme, we present the secrecy capacity of the trusted user as well as the individual achievable data rates, for a fixed number of \acp{re} of $N=40$, across a transmit \ac{snr} range of $60-120$ dB. We can observe from Fig. \ref{fig:2} that utilising '\ac{irs} only' links provides prominent secrecy capacity enhancement for medium to high \ac{snr} values. It is clear that, under low \ac{snr}, depending on the \ac{irs} will not be sufficient. The '\ac{irs} only'  strategy results in slight rate degradation for U$_t$  and a severe rate degradation for U$_u$, However, U$_u$ rate is kept higher that the minimum rate defined by the optimisation constraint in  (\ref{eq:p1c2}). Hence, the proposed scheme maximises the secrecy capacity while satisfying the rate constraints at both \ac{noma} users. We note that the \ac{los} in \ac{vlc} cannot be simply switched off so that the user only receives the reflected paths, however, we believe that the proposed results open the door for investigating novel \ac{vlc} setups in which the transmission  links are fully controlled by means of \acp{irs}. For example, by using \acp{led} with narrow \ac{fov} that can be directed towards and \ac{irs} so that the users do not perceive the \ac{los} and depend on the \ac{irs} reflections.  

\section{Conclusion} \label{sec:conclusion} 
This paper proposed a \ac{pls} mechanism for mitigating the information leakage risk  in \ac{noma} \ac{vlc} systems. The idea is based on utilising \acp{irs} to control the channel gains at the users, and hence the ability of an untrusted user to decode a confidential signal. We have demonstrated that jointly optimising the \ac{irs} configuration and \ac{noma} power allocation can enhance the secrecy capacity and reduce the possibility of eavesdropping by a potentially malicious network user. Our results revealed  that utilising \ac{irs} paths only without \ac{los} gives the highest degree of freedom and, thus, the highest possible secrecy capacity, albeit with a decrease in the achievable data rates. Although depending on \ac{irs} reflected paths without a \ac{los} link might not be ideal  from an energy efficiency perspective, we believe that the presented results offer valuable insights on how \acp{irs} can be integrated into \ac{noma}-based \ac{vlc} systems, particularly under \ac{los} blockage conditions. 

\bibliographystyle{IEEEtran}
\bibliography{Ref}

\begin{thebibliography}{10}
\providecommand{\url}[1]{#1}
\csname url@samestyle\endcsname
\providecommand{\newblock}{\relax}
\providecommand{\bibinfo}[2]{#2}
\providecommand{\BIBentrySTDinterwordspacing}{\spaceskip=0pt\relax}
\providecommand{\BIBentryALTinterwordstretchfactor}{4}
\providecommand{\BIBentryALTinterwordspacing}{\spaceskip=\fontdimen2\font plus
\BIBentryALTinterwordstretchfactor\fontdimen3\font minus
  \fontdimen4\font\relax}
\providecommand{\BIBforeignlanguage}[2]{{%
\expandafter\ifx\csname l@#1\endcsname\relax
\typeout{** WARNING: IEEEtran.bst: No hyphenation pattern has been}%
\typeout{** loaded for the language `#1'. Using the pattern for}%
\typeout{** the default language instead.}%
\else
\language=\csname l@#1\endcsname
\fi
#2}}
\providecommand{\BIBdecl}{\relax}
\BIBdecl

\bibitem{HAAS2020443}
H.~Haas, E.~Sarbazi, H.~Marshoud, and J.~Fakidis, ``Chapter 11 - visible-light
  communications and light fidelity,'' \emph{Optical Fiber Telecomm. VII}, pp.
  443--493, 2020.

\bibitem{8932632}
H.~Haas \emph{et~al.}, ``Introduction to indoor networking concepts and
  challenges in {LiFi},'' \emph{IEEE/OSA J. Opt. Commun. Netw.}, vol.~12,
  no.~2, pp. A190--A203, 2020.

\bibitem{9137669}
J.~Lian and M.~Brandt-Pearce, ``Multiuser visible light communication systems
  using {OFDMA},'' \emph{J. Lightw. Technol.}, vol.~38, no.~21, pp. 6015--6023,
  2020.

\bibitem{8713381}
H.~Abumarshoud, H.~Alshaer, and H.~Haas, ``Dynamic multiple access
  configuration in intelligent {LiFi} attocellular access points,'' \emph{IEEE
  Access}, vol.~7, pp. 62\,126--62\,141, 2019.

\bibitem{abumarshoud2021}
H.~Abumarshoud \emph{et~al.}, ``{LiFi} through reconfigurable intelligent
  surfaces: A new frontier for {6G}?'' \emph{IEEE Veh. Technol. Mag.}, pp.
  2--11, 2021.

\bibitem{9348585}
B.~{Cao} \emph{et~al.}, ``Reflecting the light: Energy efficient visible light
  communication with reconfigurable intelligent surface,'' \emph{in Proc. IEEE
  92nd Vehicular Technology Conference (VTC2020-Fall)}, pp. 1--5, Feb. 2021.

\bibitem{9526581}
S.~Sun, F.~Yang, and J.~Song, ``Sum rate maximization for intelligent
  reflecting surface-aided visible light communications,'' \emph{IEEE Commun.
  Lett.}, pp. 1--1, 2021.

\bibitem{9838853}
H.~Abumarshoud, B.~Selim, M.~Tatipamula, and H.~Haas, ``Intelligent reflecting
  surfaces for enhanced {NOMA}-based visible light communications,'' in
  \emph{ICC 2022 - IEEE International Conference on Communications}, 2022, pp.
  571--576.

\bibitem{9070153}
M.~A. Arfaoui, M.~D. Soltani, I.~Tavakkolnia, A.~Ghrayeb, M.~Safari, C.~M.
  Assi, and H.~Haas, ``Physical layer security for visible light communication
  systems: A survey,'' \emph{IEEE Commun. Surveys Tuts.}, vol.~22, no.~3, pp.
  1887--1908, 2020.

\bibitem{pls}
H.~Abumarshoud, C.~Chen, M.~S. Islim, and H.~Haas, ``Optical wireless
  communications for cyber-secure ubiquitous wireless networks,''
  \emph{Proceedings of the Royal Society A: Mathematical, Physical and
  Engineering Sciences}, vol. 476, no. 2242, Oct. 2020.

\bibitem{9524909}
H.~Abumarshoud, M.~D. Soltani, M.~Safari, and H.~Haas, ``Realistic secrecy
  performance analysis for {LiFi} systems,'' \emph{IEEE Access}, vol.~9, pp.
  120\,675--120\,688, 2021.

\bibitem{9276478}
A.~M. {Abdelhady} \emph{et~al.}, ``Visible light communications via intelligent
  reflecting surfaces: Metasurfaces vs mirror arrays,'' \emph{IEEE Open J. of
  the Commun. Soc.}, vol.~2, pp. 1--20, Dec. 2021.

\bibitem{6636053}
J.-B. Wang, Q.-S. Hu, J.~Wang, M.~Chen, and J.-Y. Wang, ``Tight bounds on
  channel capacity for dimmable visible light communications,'' \emph{J.
  Lightw. Technol.}, vol.~31, no.~23, pp. 3771--3779, 2013.

\bibitem{8421280}
J.-Y. Wang, C.~Liu, J.-B. Wang, Y.~Wu, M.~Lin, and J.~Cheng, ``Physical-layer
  security for indoor visible light communications: Secrecy capacity
  analysis,'' \emph{IEEE Trans. Commun.}, vol.~66, no.~12, pp. 6423--6436,
  2018.

\bibitem{GHANNADIAN199681}
F.~Ghannadian, C.~Alford, and R.~Shonkwiler, ``Application of random restart to
  genetic algorithms,'' \emph{Information Sciences}, vol.~95, no.~1, pp.
  81--102, 1996.

\end{thebibliography}

\end{document}